\documentclass[iop]{emulateapj}

\def\Msun{M_\odot}
\def\microas{\mu{\rm as}}

\begin{document}

\title{The PHASES Differential Astrometry Data Archive.  II.  Updated Binary Star Orbits and a Long Period Eclipsing Binary}

\author{Matthew W.~Muterspaugh\altaffilmark{1, 2}, 
William I. Hartkopf\altaffilmark{3}, 
Benjamin F.~Lane\altaffilmark{4}, 
J.~O'Connell\altaffilmark{1}, M.~Williamson\altaffilmark{2}, 
S.~R.~Kulkarni\altaffilmark{5}, 
Maciej Konacki\altaffilmark{6, 7}, Bernard F.~Burke\altaffilmark{8}, 
M.~M.~Colavita\altaffilmark{9}, M.~Shao\altaffilmark{9}, 
Sloane J.~Wiktorowicz\altaffilmark{10}}
\altaffiltext{1}{Department of Mathematics and Physics, College of Arts and 
Sciences, Tennessee State University, Boswell Science Hall, Nashville, TN 
37209 }
\altaffiltext{2}{Tennessee State University, Center of Excellence in 
Information Systems, 3500 John A. Merritt Blvd., Box No.~9501, Nashville, TN 
37209-1561}
\altaffiltext{3}{U.S.~Naval Observatory, 3450 Massachusetts Avenue, NW, Washington, DC, 20392-5420}
\altaffiltext{4}{Draper Laboratory,  555 Technology Square, Cambridge, MA 
02139-3563}
\altaffiltext{5}{Division of Physics, Mathematics and Astronomy, 105-24, 
California Institute of Technology, Pasadena, CA 91125}
\altaffiltext{6}{Nicolaus Copernicus Astronomical Center, Polish Academy of 
Sciences, Rabianska 8, 87-100 Torun, Poland}
\altaffiltext{7}{Astronomical Observatory, Adam Mickiewicz University, 
ul.~Sloneczna 36, 60-286 Poznan, Poland}
\altaffiltext{8}{MIT Kavli Institute for Astrophysics and Space Research, 
MIT Department of Physics, 70 Vassar Street, Cambridge, MA 02139}
\altaffiltext{9}{Jet Propulsion Laboratory, California Institute of 
Technology, 4800 Oak Grove Dr., Pasadena, CA 91109}
\altaffiltext{10}{Department of Astronomy, University of California, Mail Code 
3411, Berkeley, CA 94720, USA}

\email{matthew1@coe.tsuniv.edu, wih@usno.navy.mil, blane@draper.com, maciej@ncac.torun.pl}

\begin{abstract}
Differential astrometry measurements from the Palomar High-precision 
Astrometric Search for Exoplanet Systems have been combined with 
lower precision single-aperture measurements covering a much longer timespan 
(from eyepiece measurements, speckle interferometry, and adaptive optics) to 
determine improved visual orbits for 20 binary stars.  In some 
cases, radial velocity observations exist to constrain the full 
three-dimensional orbit and determine component masses.  The visual orbit of 
one of these binaries---$\alpha$ Com (HD 114378)---shows that the 
system is likely to have eclipses, despite its very long period of 
26 years.  The next eclipse is predicted to be within a week of 
2015 January 24.
\end{abstract}


\section{Introduction}

Accurate measurements of fundamental properties of stars are necessary to 
develop models of stellar formation, structure, and evolution.  These are also 
key parameters in understanding galactic mass-luminosity relationships.  
Combined visual-radial velocity (RV) orbits of binary stars provide measures 
of component masses and the distances to the systems (determining the intrinsic 
stellar luminosities), while visual-only orbits coupled with parallax 
measurements can be used to measure the total mass of the system 
\citep[see, for example][]{boden_iota_peg, 2000A&AS..145..161P, 2001AJ....121.1607B}.  
Widely separated binaries are particularly important because the stars are 
more likely to have evolved like single stars.  The 
accuracies with which the stars' fundamental properties are known are 
improved by enhanced measurement precisions.  Furthermore, establishing a 
set of high precision orbits will be useful as calibration standards for 
future precision differential astrometry efforts being pursued 
\citep{lazorenko2009, cameron2009, helminiak2009}.

A technique has been developed to obtain high precision (35 $\microas$) 
astrometry of close stellar pairs 
\citep[separation less than 1 arcsec;][]{LaneMute2004a} using 
long-baseline infrared interferometry at the Palomar Testbed 
Interferometer \citep[PTI;][]{col99}.  This technique was applied to 51 binary 
systems as the Palomar High-precision Astrometric Search for Exoplanet Systems 
(PHASES) program during 2002-2008.  PHASES science results included precision 
binary orbits and component masses, studies of the geometries and component 
physical properties in triple and quadruple star systems, and limits on the 
presence of giant planet companions to the binaries.

Astrometric measurements were made at PTI, which was 
located on Palomar Mountain near San Diego, CA. It was developed by the Jet 
Propulsion Laboratory, California Institute of Technology for NASA, as a 
testbed for interferometric techniques applicable to the Keck Interferometer 
and other missions such as the Space Interferometry Mission (SIM).  It was 
operated in the J ($1.2 \mu{\rm m}$), H ($1.6 \mu{\rm m}$), and K 
($2.2 \mu{\rm m}$) bands, and combined starlight from two out of three 
available 40-cm apertures.  The apertures formed a triangle with one 110 and 
two 87 meter baselines.  PHASES observations began in 2002 and continued 
through 2008 November when PTI ceased routine operations.

This paper is the second in a series analyzing the final results of the PHASES 
project as of its completion in late 2008.  The first paper describes the 
observing method, sources of measurement uncertainties, limits of observing 
precisions, derives empirical scaling rules to account for noise sources 
beyond those predicted by the standard reduction algorithms, and presents the 
full catalog of astrometric measurements from PHASES \citep{Mute2010A}.  The 
current paper combines PHASES astrometry with astrometric measurements made 
by other methods as well as RV observations (where available) to 
determine orbital solutions to the binaries' Keplerian motions, determining 
physical properties such as component masses and system distance when 
possible.  The third paper presents limits on the existence of substellar 
tertiary companions orbiting either the primary or secondary stars in those 
systems that are found to be consistent with being simple binaries 
\citep{Mute2010C}.  The fourth paper presents three-component orbital 
solutions to 
a known triple star system (63 Gem $=$ HD 58728) and a newly discovered 
triple system (HR 2896 $=$ HD 60318) \citep{Mute2010D}.  Finally, the fifth 
paper presents candidate substellar companions to PHASES binaries as detected 
by astrometry \citep{Mute2010E}.

\section{PHASES Measurements}

PHASES differential astrometry measurements were obtained using the observing 
method and standard data analysis pipeline described in Paper I.  The 
measurements themselves and associated measurement uncertainties are also 
tabulated in Paper I.  Analysis of binary orbits is limited to the 20 
systems for which 10 or more PHASES observations were made, do not show 
convincing evidence of having tertiary companions, and are not $\delta$ Equ, 
which has already been the subject of other PHASES investigations 
\citep{Mut05_delequ, MuteMuOri2008}.  Those having stellar companions to the 
primary and/or secondary stars either have been presented in previous works 
\citep[see][]{Mut06_kappeg, Mut06_v819her, lane88tau2007, MuteMuOri2008, lane1Gem2010_draft} 
or in Paper IV.  Those for which the PHASES measurements indicate substellar 
companions may be present are the subject of Paper V.  The number of 
PHASES measurements available for each of the 20 systems being 
investigated are listed in Table \ref{tab::speckleUnitUncert}.

\section{Non-PHASES Measurements}\label{sec::nonphases}

Measurements of binaries observed by PHASES made by previous astrometric 
techniques and cataloged in the {\em Washington Double Star Catalog} 
\citep[WDS, and see references therein, ][]{wdsCatalog, wdsCatalogUpdate} 
were assigned weights according to the formula described 
by \cite{hart01}.  Unit weight uncertainties in separation and position 
angle were evaluated by the following iterative procedure.  First, guess 
values for the unit uncertainties of 24 mas in separation and $1.8^\circ$ 
in position angle were assigned to the measurements of a given binary; these 
values corresponded to previous experience using this procedure on $\mu$ Ori 
\citep{MuteMuOri2008}.  Second, the measurements were fit to a Keplerian model 
and the orbital parameters were optimized to minimize the fit $\chi^2$.  
Third, the weighted scatter of the residuals in separation and position angle 
were evaluated, and the guessed unit uncertainties updated to make the rms 
scatter in each equal to unity.  Fourth, the second and third steps were 
iterated two more times, at which point the values converged.  Fifth, the 
final unit uncertainties were multiplied by the square root of the reduced 
$\chi^2$ ($\sqrt{\chi_r^2}$) of the fit, and refit one more time using these 
slightly larger weights.  Sixth, if no residuals deviated by more than 
3$\sigma$, the process ended, otherwise, the single measurement with the 
largest separation or 
position angle residual (weighted by its uncertainty) was flagged as an 
outlier and removed from future fits.  Seventh, the process was repeated at 
the first step.  The resulting weights are listed in Table 
\ref{tab::speckleUnitUncert} and the measurements themselves are listed in 
Table \ref{tab::speckleData}.

\begin{deluxetable}{rllrrll}
\tablecolumns{7}
\tablewidth{0pc} 
\tablecaption{Number of PHASES and Non-PHASES Measurements and Unit weight Uncertainties for Non-PHASES Measurements \label{tab::speckleUnitUncert}}
\tablehead{ 
\colhead{HD Number} & \colhead{${\rm N_{P}}$} & \colhead{${\rm N_{P, O}}$} & \colhead{${\rm N_{NP}}$} & \colhead{${\rm N_{NP, O}}$} & \colhead{$\sigma_{\rho, \circ}$} & \colhead{$\sigma_{\theta, \circ}$}}
\startdata
5286   & 18 &  0 & 645 & 34 & 0.094 & 1.82 \\
6811   & 19 &  1 & 185 & 18 & 0.031 & 2.86 \\
17904  & 45 &  1 & 150 & 20 & 0.017 & 3.95 \\
26690  & 19 &  1 &  96 & 16 & 0.0088 & 3.11 \\
44926  & 23 &  0 &  14 &  0 & 0.023 & 5.57 \\
76943  & 16 &  0 & 217 &  9 & 0.060 & 4.17 \\
77327  & 47 &  0 & 125 & 12 & 0.018 & 2.61 \\
81858  & 12 &  0 & 580 & 46 & 0.090 & 2.23 \\
114378 & 23 &  1 & 586 & 33 & 0.068 & 2.09 \\
129246 & 16 &  1 & 682 & 45 & 0.092 & 1.95 \\
137107 & 50 &  1 & 948 & 63 & 0.075 & 2.05 \\
137391 & 22 &  1 &  12 &  0 & 0.0055 & 3.25 \\
137909 & 73 &  0 &  93 & 13 & 0.016 & 2.96 \\
140159 & 15 &  0 & 205 & 13 & 0.021 & 3.78 \\
140436 & 37 &  6 & 409 & 26 & 0.078 & 3.08 \\
155103 & 10 &  0 & 131 & 23 & 0.0093 & 3.44 \\
187362 & 10 &  0 & 223 & 23 & 0.021 & 2.90 \\
202444 & 39 &  0 & 286 & 13 & 0.123 & 4.37 \\
207652 & 50 &  0 & 185 & 11 & 0.030 & 4.06 \\
214850 & 48 &  5 & 253 & 21 & 0.036 & 3.18 
\enddata
\tablecomments{
The number of PHASES and non-PHASES astrometric measurements used for orbit 
fitting with each of the binaries being studied are presented in Columns 2 and 
4 respectively, along with the additional 
numbers of measurements rejected as outliers 
in Columns 3 and 5.  Columns 6 and 7 list the 1$\sigma$ measurement 
uncertainties for unit weight measurements from 
non-PHASES observations determined by iterating Keplerian fits to the 
measurements with removal of 3$\sigma$ or greater outliers in either 
dimension.  Columns 6 and 7 are in units of arcseconds and degrees, 
respectively.  The subscripts ``P'' and ``NP'' stand for PHASES and Non-PHASES 
type measurements, respectively, while the subscript ``O'' stands for outlier 
and the subscript ``$\circ$'' represents that the uncertainties are for 
unit-weight measurements.
}
\end{deluxetable}

\begin{deluxetable*}{llllllll}
\tablecolumns{8}
\tablewidth{0pc} 
\tablecaption{Non-PHASES Astrometric Measurements \label{tab::speckleData}}
\tablehead{ 
\colhead{HD Number} & \colhead{Date} & \colhead{$\rho$} & \colhead{$\theta$} 
& \colhead{$\sigma_{\rho}$} & \colhead{$\sigma_{\theta}$} & \colhead{Weight} 
& \colhead{Outlier} \\
\colhead{} & \colhead{(yr)} & \colhead{(arcsec)} & \colhead{(deg)} 
& \colhead{(arcsec)} & \colhead{(deg)} & \colhead{} 
& \colhead{}}
\startdata
5286 & 1830.7400 & 0.850 & 305.00 & 0.297 & 5.77 & 0.1 & 0 \\
5286 & 1831.7900 & 0.780 & 308.70 & 0.297 & 5.77 & 0.1 & 0 \\
5286 & 1832.1400 & 0.850 & 307.80 & 0.086 & 1.67 & 1.2 & 0 \\
5286 & 1835.9200 & 1.100 & 315.70 & 0.297 & 5.77 & 0.1 & 0 \\
214850 & 2008.5377 & 0.363 & 77.10 & 0.008 & 0.72 & 19.4 & 0 \\
214850 & 2008.5460 & 0.364 & 77.00 & 0.008 & 0.72 & 19.4 & 0 \\
214850 & 2008.5460 & 0.364 & 77.10 & 0.008 & 0.72 & 19.4 & 0 \\
214850 & 2008.8875 & 0.380 & 77.90 & 0.024 & 2.10 & 2.3 & 0 
\enddata
\tablecomments{
Non-PHASES astrometric measurements from the WDS 
are listed with 1$\sigma$ measurements uncertainties, and weights.
Column 1 is the HD catalog number of the target star, Column 2 is 
the decimal year of the observation, Columns 3 and 4 are the separation in 
arcseconds and position angle in degrees, respectively, Columns 5 and 6 are 
the 1$\sigma$ uncertainties in the measured quantities from Columns 3 and 4, 
Column 7 is the weight assigned to the measurement, and Column 8 is 1 if the
measurement is a $>$3$\sigma$ outlier and omitted from the fit, 0 otherwise.  
(This table is available in its entirety in machine-readable and Virtual 
Observatory (VO) forms in the online journal. A portion is shown here for 
guidance regarding its form and content.)
}
\end{deluxetable*}

\section{Orbital Solutions}

\subsection{Binaries Without Radial Velocity Measurements}

Binaries for which only astrometric measurements are 
available were fit to a single Keplerian orbit using a 
downhill $\chi^2$ minimizing routine alternating 
between the standard Campbell and Thiele-Innes 
parameter sets for improved speed of convergence.  
When only a visual orbit is available, the sum of the 
component masses can be determined if a trigonometric 
parallax measurement is available.  For each system 
evaluated, the trigonometric parallax and its 
associated measurement uncertainty, as measured by 
the revised {\em Hipparcos} analysis of \cite{vanleeuwen2007}, 
were taken as input parameters to the fit in 
order to evaluate the mass sums.  When available, the 
revised parallax estimates based on the work of 
\cite[][, hereafter S99]{Soder1999} (HD 76943, HD 114378, HD 137391, 
HD 140159, HD 140436, HD 155103, 
HD 202444, and HD 207652) 
were used instead of the revised {\em Hipparcos} values because 
the S99 analysis made more complete use of non-{\em Hipparcos} 
differential astrometry values to separate the binary orbit signal 
from that of parallax.

For each of the 15 binaries, the PHASES and 
non-PHASES astrometric measurements, 
and {\em Hipparcos}-based parallax 
measurements were combined in a single orbital fit 
to evaluate the Campbell orbital parameters and system distance.
Alternatively, the orbit semimajor axis was 
replaced with the total system mass; repeating the 
fit with this substitution allowed a natural way of 
evaluating the uncertainties of the total mass.  
While the system distance is entirely dependent on the 
parallax measurement, including this as an input 
value with its associated uncertainty was necessary to 
evaluate the resulting mass uncertainties properly 
at the conclusion of the fitting procedure.  The best-fit parameters and 
their uncertainties are listed in Table \ref{tab::visualOrbits}.

\subsubsection{HD 5286}

HD 5286 (36 And, HR 258, HIP 4288, and WDS 00550$+$2338) is a pair of 
subgiant stars with spectral types G6 and K6.  The total system 
mass of $1.86 \pm 0.15 \, \Msun$ is in reasonable agreement with 
this classification \citep{allenquantities}.

\subsubsection{HD 6811}
HD 6811 ($\phi$ And, 42 And, HR 335, HIP 5434, and WDS 01095$+$4715) is 
a pair of B stars (B6IV and B9V).  The total mass is 
$6.5 \pm 2.8 \, \Msun$; while not well constrained, this value is 
reasonable for these stars.

\subsubsection{HD 17904}
HD 17904 (20 Per, HR 855, HIP 13490, and WDS 02537$+$3820) was 
reported to have a 1269 day subsystem by \cite{abt1976}.  The companion was 
further investigated by \cite{scarfe1978} and \cite{morbey1987} who found 
no evidence of such a companion, nor is such an object found 
by astrometry.  The single-component velocities from those works 
were consistent with constant velocity, providing no constraint 
on the binary orbit, and were not used for the present orbit 
fitting.  HD 17904 is a pair of mid F dwarfs; the total 
system mass of $3.88 \pm 0.58 \, \Msun$ is a bit high for such 
stars, though the uncertainty in this is large enough to still 
be in reasonable agreement.

\subsubsection{HD 44926}

While the orbital parameters of 
HD 44926 (HIP 30569, WDS 06255$+$2327) 
are not well constrained due to few 
observations covering a short fraction 
of the orbit, this is the first time an 
orbital solution has been evaluated for 
which the fit converged.

\subsubsection{HD 76943}

\cite{heintz1981} published four single-component spectroscopic 
velocities of HD 76943 (10 UMa--though it is now in the constellation Lynx 
\citep{Griffin1999}, HR 3579, HIP 44248, and WDS 09006$+$4147); the first was 
taken a year before 
the others, which were all on the same night.  More recently, 
Tennessee State University's Automatic Spectroscopic Telescope 
\citep[AST; ][]{eatonwilliamsonast} 
collected 66 blended spectra of 10 UMa spanning 1843 days 
(23 \% of the binary orbit).  Upon inspection, the single set of absorption 
lines appeared asymmetric due to the binarity, but remained blended by 
an amount substantial enough to cause systematic errors in attempts at component 
velocity determinations.  Higher resolution observations at times 
of maximal velocity separation, or individual component spectroscopy 
behind an Adaptive Optics system may allow better modeling of each 
component's lines, in which case these measurements could be recovered, but 
at this time the system remains without useful velocity measurements.

The spectral type is consistently listed as F5V, consistent 
with the total system mass of $2.44 \pm 0.12 \, \Msun$ 
\citep{allenquantities} and the S99 component mass values of 
$1.37 \pm 0.08 \, \Msun$ and $1.04 \pm 0.06 \, \Msun$; 

\subsubsection{HD 77327}

A few single component RV measurements of HD 77327 
($\kappa$ UMa, 12 UMa, HR 3594, HIP 44471, and WDS 09036$+$4709)
have been published by \cite{abt1980} and 
\cite{heintz1981}.  However, these are small in number, 
span a short time period, and appear to be consistent with constant 
velocity.  As a result, the system is treated as a visual binary only.

HD 77327 is composed of a pair of early (${\rm \sim A0}$) dwarf stars.  
The total mass of $6.30 \pm 0.98 \, {\rm \Msun}$ is not constrained 
very well, and while the value is a bit high for stars of this type, 
the large uncertainty indicates there is no cause for alarm.

\subsubsection{HD 114378}

CORAVEL produced 22 two component RV measurements of 
HD 114378 ($\alpha$ Com, 42 Com, HR 4968, HIP 64241, and WDS 13100$+$1732)
spanning nearly half (46\%) of the binary 
orbit, making it both a visual binary and double-lined spectroscopic 
binary \citep{duq91a}.  However, despite the anticipated 
velocity amplitude of each star being $\sim 8 \, {\rm km \, s^{-1}}$--several 
times larger than the $\sim 1\, {\rm km \, s^{-1}}$ typical 
measurement uncertainty--no significant velocity changes appear in 
the CORAVEL velocities.  As a result, the velocities did not agree well 
with the visual orbit (after orbit fitting, over half have residuals 
marking them as outliers of $> 4 \sigma$ compared to the published 
measurement uncertainties), and produced a system distance that was much 
too distant to agree with that of S99 ($55.9 \pm 1.4$ mas) 
based on {\em Hipparcos} astrometry.  Thus, these do not appear to 
be useful in orbit fitting.

Alternatively, \cite{tok2002} 
measured nine single-component RVs spanning $\sim 13\%$ of 
the binary orbit that were consistent with the visual orbit, though 
these are clustered at two observing epochs and showing little variation, 
making spurious orbit fitting likely.  Using these velocities 
and the parallax of $55.9 \pm 1.4$ mas from S99 yielded 
a result that is almost certainly spurious; for example, the 
binary mass ratio was a very low value (0.02).  Thus, only the visual 
orbit is presented at the current time, and additional investigations 
will need to concentrate on improving the velocity results.

The total mass was found to be $2.45 \pm 0.18 \, {\rm \Msun}$, near 
the value of $2.54 \, {\rm \Msun}$ of S99, though this 
is not surprising given that the parallax values used were identical.  
The system is a pair of F5 dwarf stars, and this mass range is 
reasonable for that spectral type.

\begin{figure}[!ht]
\epsscale{1.0}
\plotone{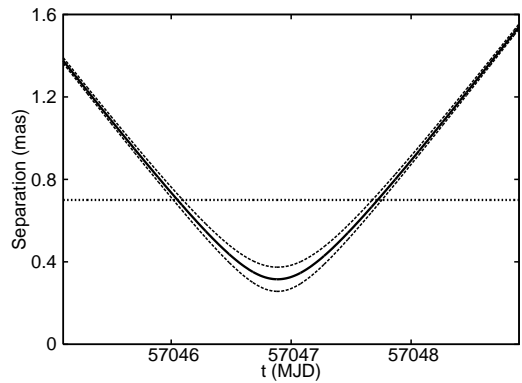}
\caption[HD 114378 ($\alpha$ Com) Eclipse Prediction]
{ \label{fig::114378_eclipse}
Time of the predicted eclipse of HD 114378 ($\alpha$ Com) in late 2015 
January.  The solid line shows the projected sky separation of the stars in 
the binary as a function of time for the best-fit orbital solution.  The 
thinner dashed lines show the separation predicted by changing the value of 
the orbital 
inclination by $\pm 1$$\sigma$.  Varying other elements of the orbital 
solution by their 1$\sigma$ uncertainties does not significantly change the 
angle of closest approach, but does change the predicted time of the eclipse.  
The horizontal line at 0.7 mas corresponds 
to the expected physical diameter of the stars.
}
\end{figure}

Most interesting of all, it is likely that this long period (26 years) 
binary eclipses!  This has been anticipated before 
\citep{Hart1989, hoffleit1996}, but previously the orbit 
was not determined well enough to demonstrate that eclipses are likely.  
The visual orbit parameters from the present 
investigation predict a closest approach of 0.32 mas around 2015 January 
24 (to within about a week), when the star is observable during early 
morning hours, and lasting about 1.5 days.  
Mid-F dwarfs have physical sizes about 1.3 times larger than the 
Sun.  At a distance of 17.9 pc, this corresponds to diameters of 0.7 
mas.  Eclipses are likely even if one considers the 1$\sigma$ upper 
limits of distance and orbital inclination, for which closest 
approach increases only to 0.38 mas and the apparent stellar 
diameters decrease by 2.5\%, an amount much smaller 
than the uncertainty in the stars' physical sizes; see 
Figure \ref{fig::114378_eclipse}.  Unfortunately, only 
one eclipse per orbit is likely for this pair; the eccentric 
orbit causes the alternative closest separation to be 0.9 mas, and a 
likely near-miss.  $\alpha$ Com, 
the brightest star in its constellation, with an orbital period 
of 26 years (nearly as long as $\epsilon$ Aur) is likely to join 
$\epsilon$ Aur as representing long period eclipsing binaries, despite 
the low probability of such a configuration.  Unlike $\epsilon$ Aur, 
in which the chance of eclipse is enhanced by the presence of an extended 
disk around one star, the eclipses in $\alpha$ Com result purely from 
orbital geometry and the overlapping of the stars themselves.

\subsubsection{HD 129246}

HD 129246 ($\zeta$ Boo, 30 Boo, HR 5477, HIP 71795, and WDS 1411$+$1344) 
has an extremely high eccentricity of $0.9977 \pm 0.0034$.  The distance of 
closest approach is only 0.3 AU.  Projected on the sky, the distance of 
closest approach is 1 mas; in this case, eclipses are unlikely, 
though the uncertainty on the inclination is large.  The next time of 
closest approach is 2023 August, though at that time the star is 
not up at night.

Though the system has a very long orbital period of 124 years, nearly 
two orbits have been observed; the first measurement was from 1796.
Despite this time coverage, and as a result of the large eccentricity, 
the semimajor axis remains poorly determined.  As a result, 
the total mass is found to be very large (122 $\Msun$) but also very 
uncertain ($\pm 275 \, \Msun$) to the point of being unconstrained.  

\subsubsection{HD 137391}

The AST attempted observations of HD 137391 
($\mu$ Boo, 51 Boo, HR 5733, HIP 75411, and WDS 15245$+$3723) several times, 
but the absorption lines were always too broad and blended for RV 
measurements.  The mass sum of $3.24 \pm 0.23 \, {\rm \Msun}$ is smaller 
than the value of 3.70 $\Msun$ from S99 
despite using the same distance to the 
system.  The current value of the semimajor axis is smaller than that 
from S99, explaining the discrepancy.

\subsubsection{HD 140159}
HD 140159 ($\iota$ Ser, 21 Set, HR 5842, HIP 76852, and WDS 15416$+$1940) 
is a pair of early A dwarfs with mass $3.73 \pm 0.53 \, \Msun$, 
consistent with both the S99 value of 3.79 $\Msun$ and 
their spectral types, though slightly on the low side for the latter 
\citep{allenquantities}.

\subsubsection{HD 140436}
HD 140436 ($\gamma$ CrB, 8 Crb, HR 5849, HIP 76952, and WDS 15427$+$2618)
is a pair of early A stars.  The total mass of $3.73 \pm 0.37 \, \Msun$ 
is slightly lower than the value of 4.23 $\Msun$ of S99, 
though still in reasonable agreement.  Like HD 140159, this mass is a 
bit low for their spectral types.

\vspace{0.5in}
\subsubsection{HD 155103}
With only 10 PHASES measurements being available, the orbit model of 
HD 155103 (c Her, HR 6377, HIP 83838, and WDS 17080$+$3556) was not much altered 
by their addition.  The total mass of $3.32 \pm 0.39 \, \Msun$ is 
identical to that from S99.  The spectral types are early (A or F), 
though the spectral class has been reported as either dwarf or giant.  
The total mass appears to be more consistent with dwarf stars.

\subsubsection{HD 187362}
HD 187362 ($\zeta$ Sge, 8 Sge, HR 7546, HIP 97496, and WDS 19490$+$1909) is 
a pair of early A dwarfs, though the measured total 
mass of $2.23 \pm 0.35 \, {\rm \Msun}$ is a bit lower than one would 
expect for such stars \citep{allenquantities}.  This may be due to a 
remaining error in the parallax:  if the original {\em Hipparcos} 
parallax is used instead, the total mass is found to be 
$4.7 \pm 1.2 \, {\rm \Msun}$, which seems more consistent \citep{hipcat}.

\subsubsection{HD 202444}

Though the binary period of 
HD 202444 ($\tau$ Cyg, 65 Cyg, HR 8130, HIP 104887, and WDS 21148$+$3803)
is long and the velocity amplitudes are low, 
it is possible to use spectroscopy to determine 
individual velocities for both components.  This is because 
one star is extremely broad lined while the 
other has very sharp spectral features.  Though the spectral lines 
are always blended, this odd pairing makes a very distinctive line profile 
that can be fit as the sum of two very different Gaussians.  
\cite{fekel2003} published two two-component RVs of the system.  
The TSU AST 
recently observed $\tau$ Cyg with high cadence to better constrain the 
mass ratio and search for short period companions to either star.  
No significant changes were observed in the velocities, and the span of 
observations was too short to contribute to orbital modeling, so at this 
time only the system's visual orbit was evaluated.

There is some indication that $\tau$ Cyg may have a substellar 
companion orbiting one of the two stars (see Paper V).  There are reasons to 
doubt the authenticity of this proposed companion, so the visual 
orbit obtained by modeling the system with only a single Keplerian model 
has been presented here in addition to the double Keplerian model presented in 
Paper V.  If real, the companion has a long orbital 
period.  This signal could be absorbed into that of the wider binary 
when only the shorter timespan PHASES data were analyzed to search for 
tertiary companions, so no compelling evidence for a companion was present 
when only PHASES measurements were analyzed.  However, the continued large 
value of $\chi^2$ 
that results when the combined PHASES and non-PHASES astrometry set described 
in Section \ref{sec::nonphases} was 
analyzed prompted a second search for tertiary companions, this time using all 
the astrometric measurements.  The longer timespan non-PHASES astrometry 
reduced the amount by which the binary orbit parameters could absorb 
motion of an intermediate period companion (short compared to the 
binary motion, long compared to the timespan of PHASES measurements) and 
indicated the presence of a companion with mass corresponding to that of 
a giant planet.

\subsubsection{HD 207652}
\begin{figure*}[!ht]
\epsscale{1.0}
\plottwo{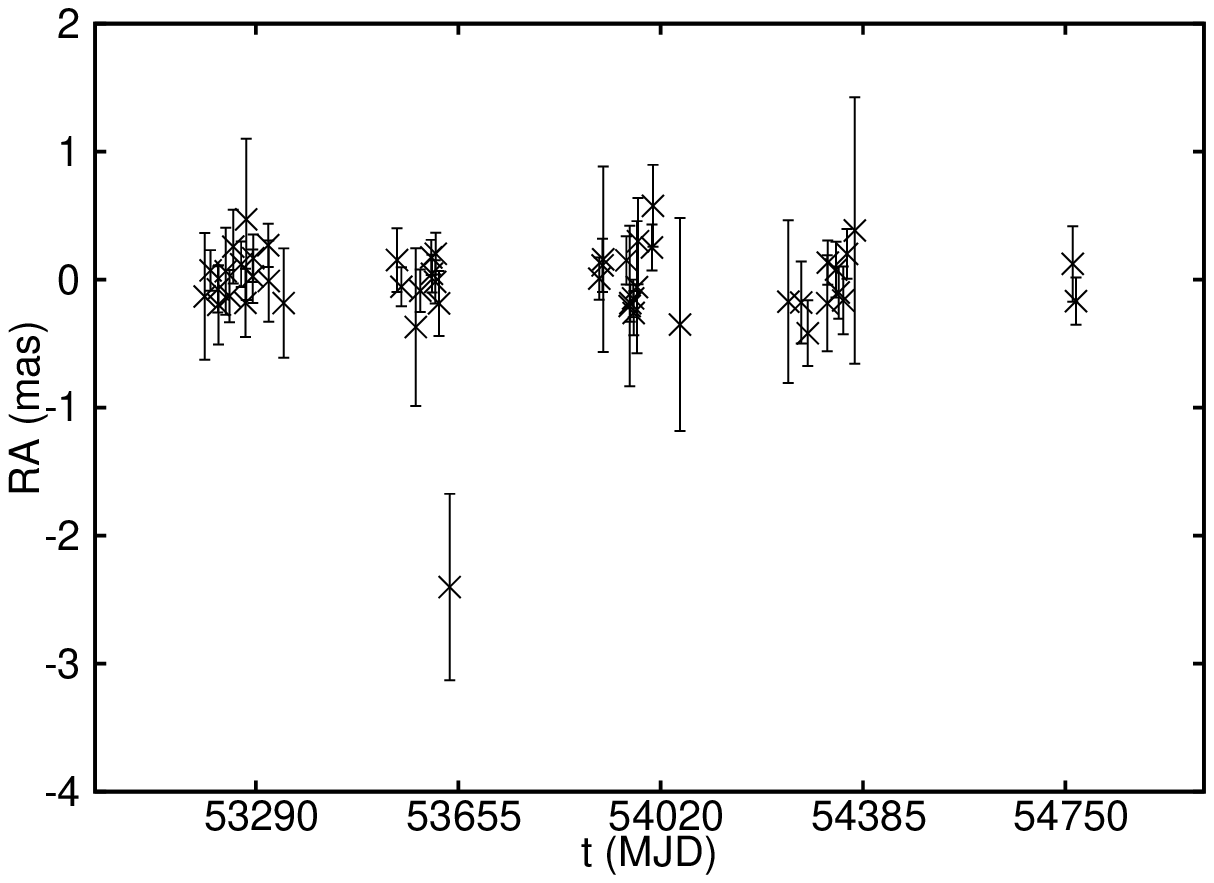}{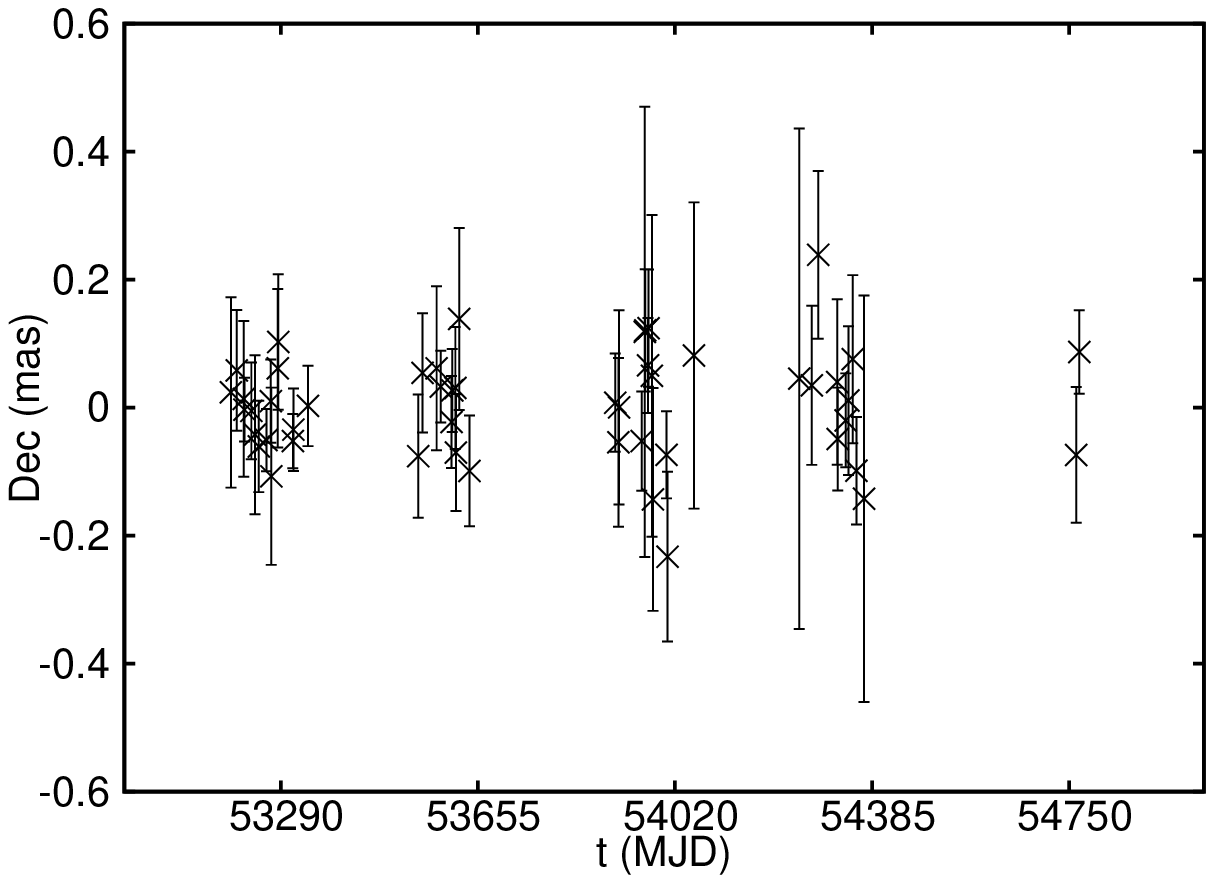}
\caption[HD 207653 (13 Peg) Fit Residuals]
{ \label{fig::207652resid}
Fit residuals to HD 207652 (13 Peg) along the differential right ascension 
(left) and differential declination (right) axes for the PHASES 
measurements.  The residuals have 
low scatter indicating that stellar variability did 
not affect the astrometry measurements.
}
\end{figure*}

HD 207652 (13 Peg, HR 8344, HIP 107788, V373 Peg, and WDS 21501$+$1717) is a 
variable flare star.  It has been suggested by \cite{1999ApJ...513..933T} that 
the secondary may be a T Tauri type variable.  
However, the variability does not appear to have affected the 
astrometry measurements; see Figure \ref{fig::207652resid}.  The total mass 
is $2.65 \pm 0.21 \, \Msun$, in 
good agreement with the value of $2.67 \, \Msun$ from S99.  For the system's 
early F spectral type, this mass and the overall system luminosity are more 
consistent with dwarf stars, though the spectral classification is giant or 
subgiant.

\subsection{Binaries With Single-component Radial Velocity Measurements}

Spectroscopic RVs can improve orbit fitting 
and enable component masses to be measured.  
If only one star's spectral features can be observed 
(single-lined spectroscopic binaries), the masses 
can be measured if the system's distance has been 
measured via trigonometric parallax and if a sufficient 
part of the orbit has been studied spectroscopically.  
Such measurements are available in the literature 
for three PHASES binaries:  46 Tau (HD 26690), 
$\omega$ Leo (HD 81858), and $\beta$ CrB (HD 137909).  Of these, 
a large number of measurements covering several 
orbital periods were available for $\beta$ CrB, while the 
others had only small amounts of data available 
spanning a fraction of the orbital period.  For $\beta$ 
CrB, including RV measurements aids constraining 
all aspects of the orbit, whereas for 46 Tau and 
$\omega$ Leo most orbital elements are unaffected, but the 
mass ratio (and thus individual masses) can be constrained 
(though at a limited level).  A separate system velocity parameter 
$V_\circ$ was used for each velocity data set of a given 
star to allow for instrumental offsets (for example, there 
are six independent $V_\circ$ parameters for $\beta$ CrB, 
one for each paper from which velocity data were obtained).  
The results of these fits are presented in Table \ref{tab::sb1Orbits}.

\subsubsection{HD 26690}

Five single-component velocity measurements covering 
488 days (18\% of the binary orbit) of HD 26690 
(46 Tau, HR 1309, HIP 19719, and WDS 04136$+$0743) were 
published by \cite{heintz1981}.  An additional 
26 unpublished measurements covering 256 days 
(10\% of the binary orbit) have been made by TSU's AST 
and are presented for the first time in Table 
\ref{tab::ast_rv_26690}.  A combined fit between the 
astrometric and velocimetry measurements found typical 
uncertainties of 1.6 ${\rm km\, s^{-1}}$ and 0.65 
${\rm km\, s^{-1}}$ were needed for these data sets, respectively, 
to make the weighted rms of the velocimetry 
residuals equal to 1.  The revised {\em Hipparcos} parallax by S99 
of $27.41 \pm 0.93$ mas was used as a weighted input in orbit fitting, 
as an observation with associated uncertainty.  The combined fit solved for 
the system distance as a parameter, and both the direct parallax measurement and orbital 
fit were used to constrain this in an optimal manner.  The component masses are 
$1.38 \pm 0.26 \, \Msun$ and $0.82 \pm 0.21 \, \Msun$, a 
mass ratio larger than that from S99, in which masses of 
$1.15 \pm 0.14 \, \Msun$ and $1.10 \pm 0.14 \, \Msun$ were found.  

\begin{deluxetable}{lll}
\tablecolumns{3}
\tablewidth{0pc} 
\tablecaption{AST Velocities of 46 Tau\label{tab::ast_rv_26690}}
\tablehead{ 
\colhead{Day (HMJD)} 
& \colhead{RV (${\rm km \, s^{-1}}$)} 
& \colhead{$\sigma_{RV} ({\rm km \, s^{-1}})$}}
\startdata
55024.46106 & -4.16 & 0.65\\
55052.38144 & -4.70 & 0.65\\
55063.44779 & -5.13 & 0.65\\
55089.51270 & -5.60 & 0.65\\
55094.51579 & -5.02 & 0.65\\
55095.30008 & -4.34 & 0.65\\
55099.50654 & -4.48 & 0.65\\
55102.41085 & -4.81 & 0.65\\
55118.43687 & -3.66 & 0.65\\
55121.46221 & -5.93 & 0.65\\
55122.46229 & -5.94 & 0.65\\
55135.38774 & -5.35 & 0.65\\
55136.26277 & -4.51 & 0.65\\
55137.28781 & -4.90 & 0.65\\
55139.37828 & -4.93 & 0.65\\
55141.42839 & -4.72 & 0.65\\
55156.33458 & -5.30 & 0.65\\
55157.11192 & -5.57 & 0.65\\
55162.26850 & -5.83 & 0.65\\
55183.28776 & -3.83 & 0.65\\
55201.39165 & -7.01 & 0.65\\
55236.11569 & -5.27 & 0.65\\
55245.16062 & -5.58 & 0.65\\
55258.19066 & -5.43 & 0.65\\
55271.17012 & -5.70 & 0.65\\
55280.11226 & -5.55 & 0.65
\enddata
\tablecomments{
Single-component RV measurements 
of 46 Tau (HD 26690) from the AST.
}
\end{deluxetable}

\vspace{0.5in}
\subsubsection{HD 81858}

\cite{abt1976} published 
20 spectroscopic measurements covering 1467 days 
(3.4\% of the binary orbit) of 
HD 81858 ($\omega$ Leo, 2 Leo, HR 3754, HIP 46454, and WDS 09285$+$0903).  
The combined astrometric and velocimetry orbital 
fit revealed that a measurement uncertainty of 1.0 
${\rm km\, s^{-1}}$ yielded a weighted rms of 1 for the 
velocity residuals.  The short span of the spectroscopic 
measurements compared to the orbital period resulted in 
only loose constraints being placed on the binary mass 
ratio and individual component masses.  The S99 parallax of 
$27.5 \pm 1.1$ mas was used as a weighted input in 
orbit fitting.  The masses are only poorly constrained as 
$1.92 \pm 0.96 \, \Msun$ and $0.28 \pm 0.92 \, \Msun$; RV measurements 
covering more of the binary orbit and/or double-line RV measurements are 
needed to improve these values.  The total system mass could be better 
determined by improved parallax measurements, but additional RV measurements 
are necessary for improved evaluation of the mass ratio.

\subsubsection{HD 137909}

A considerable number of RV measurements of HD 137909 
(``Peculiar Rosette Stone'', $\beta$ CrB, 3 CrB, HR 5747, HIP 75695, and WDS 15278$+$2906) have been 
published, spanning nearly a century.  These include the following:  
\begin{itemize}
\item 91 measurements by \cite{Cannon1912} 
($\sigma \sim 4.3 \, {\rm km\, s^{-1}}$), 
\item 352 measurements by \cite{Neubauer1944} 
($\sigma \sim 1.2 \, {\rm km\, s^{-1}}$), 
\item 25 measurements by \cite{wolff1978} 
($\sigma \sim 0.8 \, {\rm km\, s^{-1}}$), 
\item 60 measurements by \cite{Oetken1984}    
($\sigma \sim 1.0 \, {\rm km\, s^{-1}}$), 
\item 121 measurements by \cite{Kamper1990} 
(unit uncertainty $\sigma \sim 1.2 \, {\rm km\, s^{-1}}$), and 
\item 78 measurements by \cite{North1998} 
(uncertainties as given in that paper), 
\end{itemize}
where the measurement uncertainties in all but the last 
case were derived based on that which yields a weighted 
rms of 1 for the velocity residuals from a combined 
orbital fit to the astrometric and velocity data.
The S99 parallax of $29.31 \pm 0.82$ 
mas was used as a weighted input in orbit fitting.

The PHASES and non-PHASES astrometric measurements, 
velocity measurements, and S99 parallax 
measurement were combined in a single orbital fit 
to evaluate the Campbell orbital parameters as well 
as binary mass ratio and system distance.  Alternatively, 
the orbit semimajor axis and mass ratio were 
replaced with either the total system mass and mass 
ratio or the masses of each component; repeating the 
fit with these substitutions allowed a natural way of 
evaluating the uncertainties of these degenerate 
quantities.  While the system distance is entirely 
dependent on the 
parallax measurement, including this as an input 
value with its associated uncertainty was necessary to 
properly evaluate the resulting mass ratio and mass 
uncertainties at the conclusion of the fitting 
procedure.  The masses are $1.71 \pm 0.18 \, \Msun$ and 
$1.330 \pm 0.074 \, \Msun$, values a little smaller than those from 
\cite{North1998}.

\subsection{Double-lined Spectroscopic Binaries}

When the visual orbit of a binary as well as the RVs of both components are 
available, the component masses and distance to the system can be determined 
to very high precision.  The distance can be used to convert apparent 
brightnesses to absolute magnitudes, offering very strong constraints on the 
stars' physical properties.  Two PHASES binaries have RV measurements 
available for both components and the resulting orbital solutions are 
presented here.

One additional double-line spectroscopic binary 
observed with PHASES is $\delta$ Equ (HD 202275).  
Nineteen new PHASES measurements of $\delta$ Equ were 
made since its orbit was last updated in \cite{MuteMuOri2008}, 
which included 49 measurements.  However, none of the orbital elements, 
the component masses, the system distance, nor any of their associated 
uncertainties, are significantly different when all 68 
measurements were included as compared to that previous work, 
so it was not included in Table \ref{tab::sb2Orbits} because no 
updated information was needed.

\subsubsection{HD 137107}

At least one substellar companion has already been detected in the HD 137107 
($\eta$ CrB, 2 CrB, HR 5727, HIP 75312, and WDS 15232$+$3017) system, though it 
has no impact on the PHASES study.  \cite{2001AJ....121.3235K} imaged a faint 
L8 dwarf of mass $0.060 \pm 0.15 \, \Msun$ at $194''$ southeast of the A-B 
pair.  They concluded that the pair is physically associated, with a 
separation of $\sim 3600$ AU.  It is a circumbinary companion to the A-B 
binary.  

{\em CORAVEL} produced 31 two-component velocity measurements of 
HD 137107 A and B, reported first by \cite{duq91a} and revised by 
\cite{Pourbaix2000}.  The TSU AST observed this system 
33 times, but unfortunately these observations took place at a time 
when the lines were blended, and thus are not used for orbit 
fitting.  The {\em CORAVEL} measurements have been combined with 
the astrometry to develop a full three-dimensional orbit.  The S99 parallax of 
$53.5 \pm 0.9$ mas was also used as an additional measurement with 
associated uncertainty.

The masses are $1.243 \pm 0.054 \, \Msun$ and $1.100 \pm 0.039 \, \Msun$.  
As reported in Paper I, the differential magnitude of $\eta$ CrB was 
measured in the K band using Keck Adaptive Optics.  The differential 
magnitude is $\Delta K_p = 0.185 \pm 0.001$.  The apparent K magnitude from 
\cite{2MASS} is $3.714 \pm 0.216$.  Combined with the orbit-derived distance 
of $18.50 \pm 0.22$ pc (which is consistent with the trigonometric parallax), 
these correspond to absolute magnitudes of 
$K_1 = 3.04 \pm 0.22$ and $K_2 = 3.23 \pm 0.22$.  

\subsubsection{HD 214850}

\cite{batten1985} published 10 single-component velocities 
and 12 two-component velocity measurements of 
HD 214850 (HR 8631, HIP 111974, and WDS 22409$+$1433) 
spanning 1614 days ($\sim 21\%$ of the binary period) and 271 days 
($\sim 3.6\%$ of the binary period), respectively.  Four more 
measurements are omitted, as done by the original authors, due to 
partial blending of the spectral lines.  Unit uncertainties of 
$0.19 \, {\rm km \, s^{-1}}$ for component A, 
and $0.72 \, {\rm km \, s^{-1}}$ for component B 
were assigned to these measurements, resulting 
in the weighted rms scatters of each being unity after orbit fitting.  
The S99 parallax of $29.5 \pm 0.8$ mas is consistent 
with the orbital solution and has been included as an input constraint for 
orbit fitting and is reflected in the orbital parameters and their 
uncertainties in Table \ref{tab::sb2Orbits}.  Component masses are measured 
at the 7\% level.

\section{Conclusions}

The high-precision nature of the PHASES measurements and the long timespan of 
the previous astrometric measurements that aid in closing the orbits determine 
the visual orbits of many long period binaries with high accuracies.  These 
orbital solutions will benefit future astrometric programs by acting as 
calibration sources.  Future astrometric studies of these systems will aid 
the orbits presented by extending the time baseline of the precision 
observations.  Unfortunately, it is unlikely the European Gaia mission 
\citep{1996A&AS..116..579L} will contribute to the study of these specific 
systems because its detectors saturate on bright stars, though it is possible 
the proposed J-MAPS mission could do so \citep{2009IAU...261.1702G}.  
In most cases, the stellar astrophysics applications are 
limited by the RV data available.  In some cases, the lack of RV could be 
addressed simply by pursuing these measurements over the next decade or so.

\acknowledgements 
PHASES benefits from the efforts of the PTI collaboration members who have 
each contributed to the development of an extremely reliable observational 
instrument.  Without this outstanding engineering effort to produce a solid 
foundation, advanced phase-referencing techniques would not have been 
possible.  We thank PTI's night assistant Kevin Rykoski for his efforts to 
maintain PTI in excellent condition and operating PTI in phase-referencing 
mode every week.  Thanks are also extended to Ken Johnston and the 
U.~S.~Naval Observatory for their continued support of the USNO Double Star 
Program.  Part of the work described in this paper was performed at 
the Jet Propulsion Laboratory under contract with the National Aeronautics 
and Space Administration.  Interferometer data were obtained at the Palomar
Observatory with the NASA Palomar Testbed Interferometer, supported
by NASA contracts to the Jet Propulsion Laboratory.  This publication makes 
use of data products from the Two Micron All Sky Survey, which is a joint 
project of the University of Massachusetts and the Infrared Processing and 
Analysis Center/California Institute of Technology, funded by the National 
Aeronautics and Space Administration and the National Science Foundation.  
This research has made use of the Simbad database, operated at CDS, 
Strasbourg, France.  This research has made use of SAOImage DS9, developed 
by the Smithsonian Astrophysical Observatory.  
M.W.M.~acknowledges support from the Townes Fellowship 
Program, Tennessee State University, and the state of Tennessee through its 
Centers of Excellence program.  Some of the software used for analysis was 
developed as part of the SIM Double Blind Test with support from NASA 
contract NAS7-03001 (JPL 1336910).  
PHASES is funded in part by the California Institute of Technology 
Astronomy Department, and by the National Aeronautics and Space Administration 
under grant No.~NNG05GJ58G issued through the Terrestrial Planet Finder 
Foundation Science Program.  This work was supported in part by the National 
Science Foundation through grants AST 0300096, AST 0507590, and AST 0505366.
M.K.~is supported by the Foundation for Polish Science through a FOCUS 
grant and fellowship, by the Polish Ministry of Science and Higher 
Education through grant N203 3020 35.

{\it Facilities:} \facility{PO:PTI, Keck I, TSU:AST}

\bibliography{main}

\newcommand{\noopsort}[1]{} \newcommand{\printfirst}[2]{#1}
  \newcommand{\singleletter}[1]{#1} \newcommand{\switchargs}[2]{#2#1}
\begin{thebibliography}{50}
\expandafter\ifx\csname natexlab\endcsname\relax\def\natexlab#1{#1}\fi

\bibitem[{{Abt} \& {Levy}(1976)}]{abt1976}
{Abt}, H.~A. \& {Levy}, S.~G. 1976, \apjs, 30, 273

\bibitem[{{Abt} {et~al.}(1980){Abt}, {Levy}, \& {Sanwal}}]{abt1980}
{Abt}, H.~A., {Levy}, S.~G., \& {Sanwal}, N.~B. 1980, \apjs, 43, 549

\bibitem[{{Batten} {et~al.}(1985){Batten}, {Lu}, \& {Scarfe}}]{batten1985}
{Batten}, A.~H., {Lu}, W., \& {Scarfe}, C.~D. 1985, \jrasc, 79, 167

\bibitem[{{Benedict} {et~al.}(2001){Benedict}, {McArthur}, {Franz},
  {Wasserman}, {Henry}, {Takato}, {Strateva}, {Crawford}, {Ianna}, {McCarthy},
  {Nelan}, {Jefferys}, {van Altena}, {Shelus}, {Hemenway}, {Duncombe}, {Story},
  {Whipple}, {Bradley}, \& {Fredrick}}]{2001AJ....121.1607B}
{Benedict}, G.~F., {McArthur}, B.~E., {Franz}, O.~G., {Wasserman}, L.~H.,
  {Henry}, T.~J., {Takato}, T., {Strateva}, I.~V., {Crawford}, J.~L., {Ianna},
  P.~A., {McCarthy}, D.~W., {Nelan}, E., {Jefferys}, W.~H., {van Altena}, W.,
  {Shelus}, P.~J., {Hemenway}, P.~D., {Duncombe}, R.~L., {Story}, D.,
  {Whipple}, A.~L., {Bradley}, A.~J., \& {Fredrick}, L.~W. 2001, \aj, 121, 1607

\bibitem[{{Boden} {et~al.}(1999){Boden}, {Koresko}, {van Belle}, {Colavita},
  {Dumont}, {Gubler}, {Kulkarni}, {Lane}, {Mobley}, {Shao}, {Wallace}, {The PTI
  Collaboration}, \& {Henry}}]{boden_iota_peg}
{Boden}, A.~F., {Koresko}, C.~D., {van Belle}, G.~T., {Colavita}, M.~M.,
  {Dumont}, P.~J., {Gubler}, J., {Kulkarni}, S.~R., {Lane}, B.~F., {Mobley},
  D., {Shao}, M., {Wallace}, J.~K., {The PTI Collaboration}, \& {Henry}, G.~W.
  1999, \apj, 515, 356

\bibitem[{{Cameron} {et~al.}(2009){Cameron}, {Britton}, \&
  {Kulkarni}}]{cameron2009}
{Cameron}, P.~B., {Britton}, M.~C., \& {Kulkarni}, S.~R. 2009, \aj, 137, 83

\bibitem[{{Cannon}(1912)}]{Cannon1912}
{Cannon}, J.~B. 1912, \jrasc, 6, 343

\bibitem[{{Colavita} {et~al.}(1999){Colavita}, {Wallace}, {Hines}, {Gursel},
  {Malbet}, {Palmer}, {Pan}, {Shao}, {Yu}, {Boden}, {Dumont}, {Gubler},
  {Koresko}, {Kulkarni}, {Lane}, {Mobley}, \& {van Belle}}]{col99}
{Colavita}, M.~M., {Wallace}, J.~K., {Hines}, B.~E., {Gursel}, Y., {Malbet},
  F., {Palmer}, D.~L., {Pan}, X.~P., {Shao}, M., {Yu}, J.~W., {Boden}, A.~F.,
  {Dumont}, P.~J., {Gubler}, J., {Koresko}, C.~D., {Kulkarni}, S.~R., {Lane},
  B.~F., {Mobley}, D.~W., \& {van Belle}, G.~T. 1999, \apj, 510, 505

\bibitem[{{Cox}(2000)}]{allenquantities}
{Cox}, A.~N. 2000, {Allen's astrophysical quantities} (Allen's astrophysical
  quantities, 4th ed.~Publisher: New York: AIP Press; Springer, 2000.~Editedy
  by Arthur N.~Cox.~ ISBN: 0387987460)

\bibitem[{{Duquennoy} {et~al.}(1991){Duquennoy}, {Mayor}, \&
  {Halbwachs}}]{duq91a}
{Duquennoy}, A., {Mayor}, M., \& {Halbwachs}, J. 1991, \aaps, 88, 281

\bibitem[{{Eaton} \& {Williamson}(2007)}]{eatonwilliamsonast}
{Eaton}, J.~A. \& {Williamson}, M.~H. 2007, \pasp, 119, 886

\bibitem[{{Fekel} {et~al.}(2003){Fekel}, {Warner}, \& {Kaye}}]{fekel2003}
{Fekel}, F.~C., {Warner}, P.~B., \& {Kaye}, A.~B. 2003, \aj, 125, 2196

\bibitem[{{Gaume} \& {Hennessy}(2009)}]{2009IAU...261.1702G}
{Gaume}, R.~A. \& {Hennessy}, G.~S. 2009, American Astronomical Society, IAU
  Symposium \#261.~Relativity in Fundamental Astronomy: Dynamics, Reference
  Frames, and Data Analysis 27 April - 1 May 2009 Virginia Beach, VA, USA,
  \#17.02; Bulletin of the American Astronomical Society, Vol.~41, p.891, 261,
  1702

\bibitem[{{Griffin}(1999)}]{Griffin1999}
{Griffin}, R.~F. 1999, The Observatory, 119, 272

\bibitem[{Hartkopf {et~al.}(2001)Hartkopf, Mason, \& Worley}]{hart01}
Hartkopf, W.~I., Mason, B.~D., \& Worley, C.~E. 2001,
  http://www.usno.navy.mil/USNO/astrometry/optical-IR-prod/wds/orb6

\bibitem[{{Hartkopf} {et~al.}(1989){Hartkopf}, {McAlister}, \&
  {Franz}}]{Hart1989}
{Hartkopf}, W.~I., {McAlister}, H.~A., \& {Franz}, O.~G. 1989, \aj, 98, 1014

\bibitem[{{Heintz}(1981)}]{heintz1981}
{Heintz}, W.~D. 1981, \apjs, 46, 247

\bibitem[{{He{\l}miniak} {et~al.}(2009){He{\l}miniak}, {Konacki}, {Kulkarni},
  \& {Eisner}}]{helminiak2009}
{He{\l}miniak}, K.~G., {Konacki}, M., {Kulkarni}, S.~R., \& {Eisner}, J. 2009,
  \mnras, 400, 406

\bibitem[{{Hoffleit}(1996)}]{hoffleit1996}
{Hoffleit}, D. 1996, Journal of the American Association of Variable Star
  Observers (JAAVSO), 24, 105

\bibitem[{{Kamper} {et~al.}(1990){Kamper}, {McAlister}, \&
  {Hartkopf}}]{Kamper1990}
{Kamper}, K.~W., {McAlister}, H.~A., \& {Hartkopf}, W.~I. 1990, \aj, 100, 239

\bibitem[{{Kirkpatrick} {et~al.}(2001){Kirkpatrick}, {Dahn}, {Monet}, {Reid},
  {Gizis}, {Liebert}, \& {Burgasser}}]{2001AJ....121.3235K}
{Kirkpatrick}, J.~D., {Dahn}, C.~C., {Monet}, D.~G., {Reid}, I.~N., {Gizis},
  J.~E., {Liebert}, J., \& {Burgasser}, A.~J. 2001, \aj, 121, 3235

\bibitem[{{Lane} \& {Muterspaugh}(2004)}]{LaneMute2004a}
{Lane}, B.~F. \& {Muterspaugh}, M.~W. 2004, \apj, 601, 1129

\bibitem[{{Lane} {et~al.}(2007){Lane}, {Muterspaugh}, {Fekel}, {Williamson},
  {Browne}, {Konacki}, {Burke}, {Colavita}, {Kulkarni}, \&
  {Shao}}]{lane88tau2007}
{Lane}, B.~F., {Muterspaugh}, M.~W., {Fekel}, F.~C., {Williamson}, M.,
  {Browne}, S., {Konacki}, M., {Burke}, B.~F., {Colavita}, M.~M., {Kulkarni},
  S.~R., \& {Shao}, M. 2007, \apj, 669, 1209

\bibitem[{{Lane} {et~al.}(2010){Lane}, {Muterspaugh}, {Griffin}, {Scarfe},
  {Fekel}, {Williamson}, {Konacki}, {Burke}, {Colavita}, {Kulkarni}, \&
  {Shao}}]{lane1Gem2010_draft}
{Lane}, B.~F., {Muterspaugh}, M.~W., {Griffin}, R., {Scarfe}, C., {Fekel},
  F.~C., {Williamson}, M., {Konacki}, M., {Burke}, B.~F., {Colavita}, M.~M.,
  {Kulkarni}, S.~R., \& {Shao}, M. 2010, The Orbit of the Triple Star System 1
  Gem from PHASES Differential Astrometry and Radial Velocity, Submitted to
  {\apj}.

\bibitem[{{Lazorenko} {et~al.}(2009){Lazorenko}, {Mayor}, {Dominik}, {Pepe},
  {Segransan}, \& {Udry}}]{lazorenko2009}
{Lazorenko}, P.~F., {Mayor}, M., {Dominik}, M., {Pepe}, F., {Segransan}, D., \&
  {Udry}, S. 2009, \aap, 505, 903

\bibitem[{{Lindegren} \& {Perryman}(1996)}]{1996A&AS..116..579L}
{Lindegren}, L. \& {Perryman}, M.~A.~C. 1996, \aaps, 116, 579

\bibitem[{{Mason} {et~al.}(2001){Mason}, {Wycoff}, {Hartkopf}, {Douglass}, \&
  {Worley}}]{wdsCatalog}
{Mason}, B.~D., {Wycoff}, G.~L., {Hartkopf}, W.~I., {Douglass}, G.~G., \&
  {Worley}, C.~E. 2001, \aj, 122, 3466

\bibitem[{{Mason} {et~al.}(2010){Mason}, {Wycoff}, {Hartkopf}, {Douglass}, \&
  {Worley}}]{wdsCatalogUpdate}
---. 2010, http://www.usno.navy.mil/USNO/astrometry/optical-IR-prod/wds/WDS

\bibitem[{{Morbey} \& {Griffin}(1987)}]{morbey1987}
{Morbey}, C.~L. \& {Griffin}, R.~F. 1987, \apj, 317, 343

\bibitem[{{Muterspaugh} {et~al.}(2010{\natexlab{a}}){Muterspaugh}, {Fekel},
  {Lane}, {Hartkopf}, {Kulkarni}, {Konacki}, {Burke}, {Colavita}, {Shao}, \&
  {Williamson}}]{Mute2010D}
{Muterspaugh}, M.~W., {Fekel}, F.~C., {Lane}, B.~F., {Hartkopf}, W.~I.,
  {Kulkarni}, S.~R., {Konacki}, M., {Burke}, B.~F., {Colavita}, M.~M., {Shao},
  M., \& {Williamson}, M. 2010{\natexlab{a}}, Submitted to \aj

\bibitem[{{Muterspaugh} {et~al.}(2008){Muterspaugh}, {Lane}, {Fekel},
  {Konacki}, {Burke}, {Kulkarni}, {Colavita}, {Shao}, \&
  {Wiktorowicz}}]{MuteMuOri2008}
{Muterspaugh}, M.~W., {Lane}, B.~F., {Fekel}, F.~C., {Konacki}, M., {Burke},
  B.~F., {Kulkarni}, S.~R., {Colavita}, M.~M., {Shao}, M., \& {Wiktorowicz},
  S.~J. 2008, \aj, 135, 766

\bibitem[{{Muterspaugh} {et~al.}(2005){Muterspaugh}, {Lane}, {Konacki},
  {Burke}, {Colavita}, {Kulkarni}, \& {Shao}}]{Mut05_delequ}
{Muterspaugh}, M.~W., {Lane}, B.~F., {Konacki}, M., {Burke}, B.~F., {Colavita},
  M.~M., {Kulkarni}, S.~R., \& {Shao}, M. 2005, \aj, 130, 2866

\bibitem[{{Muterspaugh} {et~al.}(2006{\natexlab{a}}){Muterspaugh}, {Lane},
  {Konacki}, {Burke}, {Colavita}, {Kulkarni}, \& {Shao}}]{Mut06_v819her}
---. 2006{\natexlab{a}}, \aap, 446, 723

\bibitem[{{Muterspaugh} {et~al.}(2006{\natexlab{b}}){Muterspaugh}, {Lane},
  {Konacki}, {Wiktorowicz}, {Burke}, {Colavita}, {Kulkarni}, \&
  {Shao}}]{Mut06_kappeg}
{Muterspaugh}, M.~W., {Lane}, B.~F., {Konacki}, M., {Wiktorowicz}, S., {Burke},
  B.~F., {Colavita}, M.~M., {Kulkarni}, S.~R., \& {Shao}, M.
  2006{\natexlab{b}}, \apj, 636, 1020

\bibitem[{{Muterspaugh} {et~al.}(2010{\natexlab{b}}){Muterspaugh}, {Lane},
  {Kulkarni}, {Konacki}, {Burke}, {Colavita}, \& {Shao}}]{Mute2010C}
{Muterspaugh}, M.~W., {Lane}, B.~F., {Kulkarni}, S.~R., {Konacki}, M., {Burke},
  B.~F., {Colavita}, M.~M., \& {Shao}, M. 2010{\natexlab{b}}, Submitted to \aj

\bibitem[{{Muterspaugh} {et~al.}(2010{\natexlab{c}}){Muterspaugh}, {Lane},
  {Kulkarni}, {Konacki}, {Burke}, {Colavita}, {Shao}, {Hartkopf}, {Boss}, \&
  {Williamson}}]{Mute2010E}
{Muterspaugh}, M.~W., {Lane}, B.~F., {Kulkarni}, S.~R., {Konacki}, M., {Burke},
  B.~F., {Colavita}, M.~M., {Shao}, M., {Hartkopf}, W.~I., {Boss}, A.~P., \&
  {Williamson}, M. 2010{\natexlab{c}}, Submitted to \aj

\bibitem[{{Muterspaugh} {et~al.}(2010{\natexlab{d}}){Muterspaugh}, {Lane},
  {Kulkarni}, {Konacki}, {Burke}, {Colavita}, {Shao}, {Wiktorowicz}, \&
  {O'Connell}}]{Mute2010A}
{Muterspaugh}, M.~W., {Lane}, B.~F., {Kulkarni}, S.~R., {Konacki}, M., {Burke},
  B.~F., {Colavita}, M.~M., {Shao}, M., {Wiktorowicz}, S.~J., \& {O'Connell},
  J. 2010{\natexlab{d}}, Submitted to \aj

\bibitem[{{Neubauer}(1944)}]{Neubauer1944}
{Neubauer}, F.~J. 1944, \apj, 99, 134

\bibitem[{{North} {et~al.}(1998){North}, {Carquillat}, {Ginestet}, {Carrier},
  \& {Udry}}]{North1998}
{North}, P., {Carquillat}, J.-M., {Ginestet}, N., {Carrier}, F., \& {Udry}, S.
  1998, \aaps, 130, 223

\bibitem[{{Oetken} \& {Orwert}(1984)}]{Oetken1984}
{Oetken}, L. \& {Orwert}, R. 1984, Astronomische Nachrichten, 305, 317

\bibitem[{{Perryman} {et~al.}(1997){Perryman}, {Lindegren}, {Kovalevsky},
  {Hoeg}, {Bastian}, {Bernacca}, {Cr{\' e}z{\' e}}, {Donati}, {Grenon}, {van
  Leeuwen}, {van der Marel}, {Mignard}, {Murray}, {Le Poole}, {Schrijver},
  {Turon}, {Arenou}, {Froeschl{\' e}}, \& {Petersen}}]{hipcat}
{Perryman}, M.~A.~C., {Lindegren}, L., {Kovalevsky}, J., {Hoeg}, E., {Bastian},
  U., {Bernacca}, P.~L., {Cr{\' e}z{\' e}}, M., {Donati}, F., {Grenon}, M.,
  {van Leeuwen}, F., {van der Marel}, H., {Mignard}, F., {Murray}, C.~A., {Le
  Poole}, R.~S., {Schrijver}, H., {Turon}, C., {Arenou}, F., {Froeschl{\' e}},
  M., \& {Petersen}, C.~S. 1997, \aap, 323, L49

\bibitem[{{Pourbaix}(2000)}]{Pourbaix2000}
{Pourbaix}, D. 2000, \aaps, 145, 215

\bibitem[{{Pourbaix} \& {Jorissen}(2000)}]{2000A&AS..145..161P}
{Pourbaix}, D. \& {Jorissen}, A. 2000, \aaps, 145, 161

\bibitem[{{Scarfe} \& {Fekel}(1978)}]{scarfe1978}
{Scarfe}, C.~D. \& {Fekel}, F.~C. 1978, \pasp, 90, 297

\bibitem[{{Skrutskie} {et~al.}(2006){Skrutskie}, {Cutri}, {Stiening},
  {Weinberg}, {Schneider}, {Carpenter}, {Beichman}, {Capps}, {Chester},
  {Elias}, {Huchra}, {Liebert}, {Lonsdale}, {Monet}, {Price}, {Seitzer},
  {Jarrett}, {Kirkpatrick}, {Gizis}, {Howard}, {Evans}, {Fowler}, {Fullmer},
  {Hurt}, {Light}, {Kopan}, {Marsh}, {McCallon}, {Tam}, {Van Dyk}, \&
  {Wheelock}}]{2MASS}
{Skrutskie}, M.~F., {Cutri}, R.~M., {Stiening}, R., {Weinberg}, M.~D.,
  {Schneider}, S., {Carpenter}, J.~M., {Beichman}, C., {Capps}, R., {Chester},
  T., {Elias}, J., {Huchra}, J., {Liebert}, J., {Lonsdale}, C., {Monet}, D.~G.,
  {Price}, S., {Seitzer}, P., {Jarrett}, T., {Kirkpatrick}, J.~D., {Gizis},
  J.~E., {Howard}, E., {Evans}, T., {Fowler}, J., {Fullmer}, L., {Hurt}, R.,
  {Light}, R., {Kopan}, E.~L., {Marsh}, K.~A., {McCallon}, H.~L., {Tam}, R.,
  {Van Dyk}, S., \& {Wheelock}, S. 2006, \aj, 131, 1163

\bibitem[{S{\" o}derhjelm(1999)}]{Soder1999}
S{\" o}derhjelm, S. 1999, A\&A, 341, 121

\bibitem[{{Tamazian} {et~al.}(1999){Tamazian}, {Docobo}, \&
  {Melikian}}]{1999ApJ...513..933T}
{Tamazian}, V.~S., {Docobo}, J.~A., \& {Melikian}, N.~D. 1999, \apj, 513, 933

\bibitem[{{Tokovinin} \& {Smekhov}(2002)}]{tok2002}
{Tokovinin}, A.~A. \& {Smekhov}, M.~G. 2002, \aap, 382, 118

\bibitem[{{van Leeuwen}(2007)}]{vanleeuwen2007}
{van Leeuwen}, F., ed. 2007, Astrophysics and Space Science Library, Vol. 350,
  {Hipparcos, the New Reduction of the Raw Data}

\bibitem[{{Wolff}(1978)}]{wolff1978}
{Wolff}, S.~C. 1978, \pasp, 90, 412

\end{thebibliography}
\bibliographystyle{apj}

\clearpage
\LongTables 
\begin{deluxetable*}{lccccccccc}
\tablecolumns{10}
\tablewidth{0pc} 
\tablecaption{Visual Orbit Parameters\label{tab::visualOrbits}}
\tablehead{ 
\colhead{HD Number} 
& \colhead{Period (d)} 
& \colhead{${\rm T_\circ}$ (HMJD)} 
& \colhead{Semimajor Axis (arcsec)} 
& \colhead{Eccentricity} 
& \colhead{Inclination (deg)} 
& \colhead{$\omega$ (deg)} 
& \colhead{$\Omega$ (deg)}
& \colhead{$M_1+M_2$ ($\Msun$)}
& \colhead{$\chi^2$}\\
\colhead{} 
& \colhead{$\sigma_{\rm P}$} 
& \colhead{$\sigma_{\rm T_\circ}$} 
& \colhead{$\sigma_a$} 
& \colhead{$\sigma_e$} 
& \colhead{$\sigma_i$} 
& \colhead{$\sigma_\omega$} 
& \colhead{$\sigma_\Omega$}
& \colhead{$\sigma_{M_1+M_2}$}
& \colhead{dof}}
\startdata
5286   &   61183     & 35543     & 0.9837    & 0.30603  &  44.57  & 358.62   & 173.66   &  1.86  & 1574.1 \\
       &     (69)    &   (21)   & (0.0011)  & (0.00078) &  (0.11) &  (0.21)  &  (0.13)  & (0.15) & (1319) \\  
6811   &  202459     & 17740     & 0.573     & 0.385    & 142.2   & 112.6    & 337.2    &  6.5   &  464.1 \\
       &  (24530)    & (1837)    & (0.051)   & (0.043)  &  (2.8)  &  (9.1)   &  (3.2)   & (2.8)  & (401)  \\
17904  &   11553.9   & 50255     & 0.2224    & 0.7560   & 120.48  & 265.54   &  26.62   &  3.88  &  360.7 \\
       &      (8.7)  &   (12)    & (0.0011)  & (0.0023) &  (0.20) &  (0.11)  &  (0.24)  & (0.58) & (383)  \\
44926  &  339008     & 53508     & 0.5       & 0.7      &  80.7   &  14      & 342.18   &  2.53  &   43.1 \\
       & (3132095)   & (2553)    & (3.0)     & (1.6)    &  (4.9)  & (23)     &  (0.70)  & (1.3)   &  (67)  \\
76943  &  7691.0     & 49262.6   &  0.64566  &  0.15075  & 131.366  & 32.30  & 203.74   &  2.44  &  476.9 \\
       &    (1.8)    &    (9.1)  & (0.00065) & (0.00084) &  (0.099) & (0.44) &  (0.10)  & (0.12) &  (459) \\
77327  &   13007.2   & 50404     & 0.18194   & 0.5584   & 109.410 &  355.63  & 105.641  & 6.30   &  297.0 \\
       &      (9.7)  &   (12)    & (0.00025) & (0.0015) & (0.066) &   (0.36) &  (0.080) & (0.98) &  337   \\
114378 &    9485.68  & 47651.8   & 0.66132   & 0.4957   &  90.054 &  101.689 &  12.221  &  2.45  & 1162.3 \\
       &      (0.97) &    (2.6)  & (0.00061) & (0.0010) & (0.010) &  (0.059) &  (0.015) & (0.18) & (1211) \\
129246 &   45460     & 60183     & 2.3      & 0.9977   & 102.3   &  262.9   &   8.2    & 122    & 1452.2 \\
       &     (62)    &   (57)    & (1.7)    & (0.0034) &  (9.2)  &   (5.9)  &  (2.6)   & (275)  & (1389) \\
137391 &    1368.02  & 53855.92  & 0.098837  & 0.27194  & 130.016 &   44.204 & 130.040  &  3.24  &   52.1 \\
       &      (0.24) &    (0.23) & (0.000034) & (0.00028) & (0.029) & (0.068) & (0.040) & (0.23) &  (61)  \\
140159 &    8015.0   & 54180     & 0.21033   &  0.0941  &  83.608 &   80.5   &  69.684  &  3.73  &  453.1 \\
       &      (6.3)  &   (33)    & (0.00047) & (0.0028) & (0.043) &   (1.8)  &  (0.033) & (0.53) & (433)  \\
140436 &   33701     & 59947     &  0.7315   &  0.4779  &  94.263 &  283.40  & 292.043  &  3.73  & 1017.4 \\
       &     (61)    &   (63)    & (0.0034)  & (0.0067) & (0.028) &   (0.36) &  (0.036) & (0.37) & (885)  \\
155103 &    2971.54  & 42567.9   & 0.10863   & 0.5338   & 121.221 &   53.81  & 309.30   &  3.32  &  274.3 \\
       &      (0.84) &    (3.4)  & (0.00020) & (0.0015) & (0.077) &   (0.16) &  (0.11)  & (0.39) & (275)  \\
187362 &    8487.9   & 44199.6   & 0.13605   & 0.7948   & 132.33  &  355.3   & 340.97   &  2.23   &  467.3 \\
       &      (4.9)  &    (5.9)  & (0.00044) & (0.0019) &  (0.41) &   (1.0)  &  (0.68)  & (0.35)  & (459)  \\
202444 &    18122.6  & 47527     & 0.9160    & 0.2413   & 134.11  &  298.28  & 339.72   &  2.63  &  994.7 \\
       &       (8.8) &   (18)    & (0.0013)  & (0.0013) &  (0.14) &   (0.20) &  (0.14)  & (0.12) & (643)  \\
207652 &     9622.8  & 47848.3   & 0.36245   & 0.2300   &  70.379 &  250.23  & 230.698  &  2.65  &  431.8 \\
       &       (4.1) &    (6.2)  & (0.00026) & (0.0011) & (0.029) &   (0.14) &  (0.028) & (0.21) & (463)  
\enddata
\tablecomments{
The model parameters and fit uncertainties derived from a simultaneous fit to 
PHASES and non-PHASES measurements.
}
\end{deluxetable*}

\begin{deluxetable*}{lcccccccccccc}
\tablecolumns{13}
\tablewidth{0pc} 
\tablecaption{Single-line Spectroscopic Binary Orbit Parameters\label{tab::sb1Orbits}}
\tablehead{ 
\colhead{HD Number} 
& \colhead{Period (d)} 
& \colhead{${\rm T_\circ}$ (HMJD)} 
& \colhead{Semimajor Axis (arcsec)} 
& \colhead{Eccentricity} 
& \colhead{Inclination (deg)} 
& \colhead{$\omega$ (deg)} 
& \colhead{$\Omega$ (deg)}
& \colhead{$M_2/M_1$}
& \colhead{$M_1+M_2$ ($\Msun$)}
& \colhead{$M_1$ ($\Msun$)}
& \colhead{$M_2$ ($\Msun$)}
& \colhead{$\chi^2$}\\
\colhead{} 
& \colhead{$\sigma_{\rm P}$} 
& \colhead{$\sigma_{\rm T_\circ}$} 
& \colhead{$\sigma_a$} 
& \colhead{$\sigma_e$} 
& \colhead{$\sigma_i$} 
& \colhead{$\sigma_\omega$} 
& \colhead{$\sigma_\Omega$}
& \colhead{$\sigma_{M_1/M_2}$}
& \colhead{$\sigma_{M_1+M_2}$}
& \colhead{$\sigma_{M_1}$}
& \colhead{$\sigma_{M_2}$}
& \colhead{DOF}}
\startdata
26690  &    2633.34  & 53415.37  & 0.13302   & 0.3301   &  66.942 & 306.658  & 144.296  & 0.60 & 2.20 & 1.38 & 0.82 & 264.2 \\
       &      (0.50) &    (0.39) & (0.00015) & (0.0010) & (0.039) &  (0.071) &  (0.040) & (0.23) & (0.23) & (0.26) & (0.21) & (251)   \\
81858  &   43089     & 36649     & 0.8599    & 0.5619   &  65.347 &  302.54  & 325.982  & 0.14 & 2.20 & 1.92 & 0.28 & 1236.4 \\
       &     (26)    &   (11)    & (0.0022)  & (0.0014) & (0.093) &   (0.14) &  (0.074) & (0.55) & (0.27) & (0.96) & (0.92) & (1195)   \\
137909 &    3848.54  & 44412.8   & 0.204008  & 0.53971  & 111.452 &  180.21  & 148.041  & 0.779 & 3.04 & 1.71 & 1.330 & 1023.5 \\
       &      (0.52) &    (1.4)  & (0.000034) & (0.00021) & (0.014) & (0.13) &  (0.030) & (0.039) & (0.25) & (0.18) & (0.074) & (1045)   
\enddata
\tablecomments{
The model parameters and fit uncertainties derived from a simultaneous fit to 
PHASES and non-PHASES measurements.
}
\end{deluxetable*}

\begin{deluxetable*}{lccccccccccccc}
\tablecolumns{14}
\tablewidth{0pc} 
\tablecaption{Double-line Spectroscopic Binary Orbit Parameters\label{tab::sb2Orbits}}
\tablehead{ 
\colhead{HD Number} 
& \colhead{Period (d)} 
& \colhead{${\rm T_\circ}$ (HMJD)} 
& \colhead{Semimajor Axis (arcsec)} 
& \colhead{Eccentricity} 
& \colhead{Inclination (deg)} 
& \colhead{$\omega$ (deg)} 
& \colhead{$\Omega$ (deg)}
& \colhead{$M_2/M_1$}
& \colhead{$M_1+M_2$ ($\Msun$)}
& \colhead{$M_1$ ($\Msun$)}
& \colhead{$M_2$ ($\Msun$)}
& \colhead{d (parsec)}
& \colhead{$\chi^2$}\\
\colhead{} 
& \colhead{$\sigma_{\rm P}$} 
& \colhead{$\sigma_{\rm T_\circ}$} 
& \colhead{$\sigma_a$} 
& \colhead{$\sigma_e$} 
& \colhead{$\sigma_i$} 
& \colhead{$\sigma_\omega$} 
& \colhead{$\sigma_\Omega$}
& \colhead{$\sigma_{M_1/M_2}$}
& \colhead{$\sigma_{M_1+M_2}$}
& \colhead{$\sigma_{M_1}$}
& \colhead{$\sigma_{M_2}$}
& \colhead{$\sigma_d$}
& \colhead{DOF}}
\startdata
137107 &   15204.9   & 42612.9   & 0.86226   & 0.27907  &  58.084 &   39.885 & 202.827  & 0.885 & 2.343 & 1.243 & 1.100 & 18.50 & 2102.6 \\
       &      (1.4)  &    (3.4)  & (0.00033) & (0.00026) & (0.026) & (0.064) &  (0.024) & (0.031) & (0.084) & (0.054) & (0.039) & (0.22) & (2049) \\
214850 &    7607.7   & 45531.7   & 0.287980  & 0.73499  & 139.861 &   22.31  & 251.540  & 1.089 & 2.246 & 1.075 & 1.171 & 34.43 & 633.4 \\
       &      (1.1)  &    (1.2)  & (0.000049) & (0.00014) & (0.032) & (0.12) &  (0.076) & (0.080) & (0.067) & (0.058) & (0.047) & (0.34) & (627) 
\enddata
\tablecomments{
The model parameters and fit uncertainties derived from a simultaneous fit to 
PHASES and non-PHASES astrometry and two component RV measurements.
}
\end{deluxetable*}


\end{document}